\documentclass{article}

\usepackage{arxiv}

\usepackage[utf8]{inputenc} 
\usepackage[T1]{fontenc}    
\usepackage{hyperref}       
\usepackage{url}            
\usepackage{booktabs}       
\usepackage{amsfonts}       
\usepackage{nicefrac}       
\usepackage{microtype}      
\usepackage{lipsum}		
\usepackage{graphicx}
\usepackage{natbib}
\usepackage{doi}

\usepackage{tikz}
\usetikzlibrary{shapes.geometric, arrows}
\tikzstyle{startstop} = [rectangle, rounded corners, minimum width=3cm, minimum height=1cm,text centered, draw=black, fill=red!30]
\tikzstyle{io} = [trapezium, trapezium left angle=70, trapezium right angle=110, minimum width=3cm, minimum height=1cm, text centered, draw=black, fill=blue!30]
\tikzstyle{process} = [rectangle, minimum width=3cm, minimum height=1cm, text centered, draw=black, fill=orange!30]
\tikzstyle{decision} = [diamond, minimum width=3cm, minimum height=1cm, text centered, draw=black, fill=green!30]
\tikzstyle{arrow} = [thick,->,>=stealth]
\usepackage{booktabs}
\usepackage{chronology}
\usepackage{float}

\usepackage{multicol}
\usepackage{enumitem}
\usepackage{outlines}

\newcommand{\newCommandName}{Draft 3.2}

\usepackage{sectsty}
\subsubsectionfont{\normalfont\itshape}

\usepackage{titlesec}
\titlespacing*{\section}{0pt}{1.1\baselineskip}{\baselineskip}
\titlespacing{\section}{0pt}{1.1\baselineskip}{\baselineskip}
\usepackage{lineno}

\title{Baseline validation of a bias-mitigated loan screening model based on the European Banking Authority's trust elements of Big Data \& Advanced Analytics applications using Artificial Intelligence}

\author{ \href{https://orcid.org/0000-0003-2272-132X}{\includegraphics[scale=0.06]{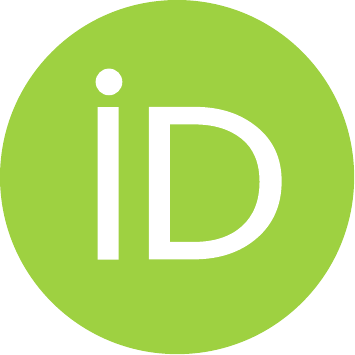}\hspace{1mm}Alessandro Danovi}\\
	University of Bergamo, Bergamo 24100, Italy \\
	\texttt{alessandro.danovi@unibg.it} \\
	\And
	Marzio Roma \\
	Alba Leasing, Milano 20139, Italy \\
	\texttt{marzio.roma@albaleasing.eu} \\
	\And
	\href{https://orcid.org/0000-0003-0138-2461}{\includegraphics[scale=0.06]{orcid.pdf}\hspace{1mm}Davide Meloni} \\
	QuantumSPEKTRAL,  Milano 20100, Italy \\
	\texttt{davide@quantumspektral.com} \\
	\And
	\href{https://orcid.org/0000-0001-5181-8433}{\includegraphics[scale=0.06]{orcid.pdf}\hspace{1mm}Stefano Olgiati *} \\
	University of Ferrara, Ferrara 44121, Italy \\
	 ~* \texttt{stefano.olgiati@unife.it} \\
	\And
	Fernando Metelli \\
	Alba Leasing, Milano 20139, Italy \\
	\texttt{fernando.metelli@albaleasing.eu} 
	}

\date{}

\renewcommand{\headeright}{\newCommandName}
\renewcommand{\undertitle}{Research Report Preprint}
\renewcommand{\shorttitle}{ \textit{Baseline validation of a bias-mitigated loan screening model based on the European Banking Authority's trust elements \\ of Big Data \& Advanced Analytics applications using Artificial Intelligence}}

\hypersetup{
pdftitle={Baseline validation of a bias-mitigated loan screening model based on the European Banking Authority's trust elements of Big Data \& Advanced Analytics applications using Artificial Intelligence},
pdfsubject={q-fin.RM},
pdfauthor={Alessandro Danovi, Marzio Roma, Davide Meloni, Stefano Olgiati, Fernando Metelli},
pdfkeywords={Loan screening, Bias mitigation, Big Data and Advanced Analytics, Artificial Intelligence   },
}

\begin{document}
\maketitle
\vspace{-0.6cm}
\begin{abstract}
\textbf{Background} 
The goal of our 4-phase research project was to test if a machine-learning-based loan screening application (5D) could detect bad loans: 1) in a subset of non-performing loans, and; 2) in a set of performing and non-performing loans, subject to the following constraints: a) utilize a minimal-optimal number of publicly available features unrelated to the credit history, gender, race or ethnicity of the borrower (BiMOPT features); b) comply with the European Banking Authority and EU Commission AI HLEG principles on trustworthy Artificial Intelligence (AI). \textbf{Methods} All datasets have been anonymized and pseudoanonymized. In Phase 0 we selected a subset of 10 BiMOPT features out of a total of 84 features; in Phase I we trained 5D to detect bad loans in a historical dataset extracted from a mandatory report to the Bank of Italy consisting of 7,289 non-performing loans (NPLs) closed in the period 2010-2021; in Phase II we assessed the baseline performance of 5D on a distinct validation dataset consisting of an active portolio of 63,763 outstanding loans (performing and non-performing) for a total financed value of over EUR 11.5 billion as of December 31, 2021;  in Phase III we will monitor the baseline performance for a period of 5 years (2023-27) to assess the prospective real-world bias-mitigation and performance of the 5D system and its utility in credit and fintech institutions \textbf{Results} 5D correctly detected 1,461 bad loans out of a total of 1,613 (Sensitivity = 0.91, Prevalence = 0.0253;, Positive Predictive Value = 0.19), and correctly classified 55,866 out of the other 62,150 exposures (Specificity = 0.90,  Negative Predictive Value = 0.997). \textbf{Interpretation} Our preliminary results support the hypothesis that Big Data \& Advanced Analytics applications based on AI can mitigate bias and improve consumer protection in the loan screening process without compromising the efficacy of the credit risk assessment. Further validation is required to assess the prospective performance and utility of 5D in credit and fintech institutions. \textbf{Funding} QuantumSPEKTRAL and Alba Leasing.
\end{abstract}

\keywords{Loan screening \and Bias mitigation \and Big Data and Advanced Analytics \and Artificial Intelligence   }


\newpage
\begin{multicols}{2}

\section{Background}
\label{sec:Background}

\subsection{The European Banking Authority Report on Big Data \& Advanced Analytics applications} 
The European Banking Authority (EBA) published in 2020 a report on  big data and advanced analytics (BD\&AA) where it observed a fast-growing interest in the use BD\&AA solutions, with two out of three EU credit institutions already deploying such solutions \citenum{EBA_2020}. \\
EBA identified: i) the four functions of BD\&AA applications; ii) the four key pillars of BD\&AA applications, and; iii) the eleven fundamental trust elements of BD\&AA applications based on Artificial Intelligence.

\subsubsection{The four functions of BD\&AA applications} 
EBA observed that all functions across institutions could benefit from BD\&AA applications, as they can improve  \citenum{EBA_2016},:
\begin{enumerate}[leftmargin=3mm, align=left,labelwidth=\parindent,labelsep=2pt]
\item efficiency;
\item productivity;
\item the cost-effectiveness of existing services, and;
\item create new business opportunities.
\end{enumerate}
EBA remarked that presently EU credit institutions are using software applications based on simpler algorithms than BD\&AA, limiting themselves to leveraging on their existing core data but that the current landscape can evolve at a rapid pace in the next few years. \\
This evolution introduces the necessity to define the key pillars and fundamental trust elements upon which BD\&AA applications should be built.

\subsubsection{The four key pillars of BD\&AA applications}
The EBA report identifies the four key pillars for the development, implementation, and adoption of BD\&AA  \citenum{EBA_2018}:
\begin{enumerate}[leftmargin=3mm,align=left,labelwidth=\parindent,labelsep=2pt]
\item  data management;
\item technological infrastructure;
\item analytics methodology, and;
\item organization and governance.
\end{enumerate}

\subsubsection{The eleven fundamental trust elements of BD\&AA applications based on Artificial Intelligence}
EBA finds that the roll-out of BD\&AA applications based on artificial intelligence/machine learning is based on the principle of trustworthiness, and the European Commission notes that all  BD\&AA applications deployed by credit institutions in the EU should respect these eleven fundamental trust elements and that they have implications for all the four key pillars. 
These eleven trust elements are \citenum{EU_2019_1}:
\begin{enumerate}[leftmargin=3mm,align=left,labelwidth=\parindent,labelsep=2pt]
\item ethics;
\item explainability;
\item interpretability;
\item fairness;
\item avoidance of bias;
\item traceability;
\item auditability;
\item data protection;
\item data quality;
\item security;
\item consumer protection.
\end{enumerate}
 
 \subsection{The position of the European Commission High-Level Expert Group on Artificial Intelligence }
The  European Commission High-Level Expert Group on Artificial Intelligence (AI HLEG) has also prepared a document which elaborates the Guidelines on trustworthy AI used. Machine Learning is a subcategory of AI.

\subsubsection{The three principles of trustworthy Artificial Intelligence}
According to the AI HLEG Guidelines, trustworthy AI should be \citenum{EU_2019_1}:
\begin{enumerate}[leftmargin=3mm,align=left,labelwidth=\parindent,labelsep=2pt]
\item lawful -  respecting all applicable laws and regulations;
\item ethical - respecting ethical principles and values;
\item robust - both from a technical perspective while taking into account its social environment.
\end{enumerate}

\subsubsection{The seven key requirements of Artificial Intelligence systems}
The Guidelines put forward by AI HLEG are a set of seven key requirements that AI systems should meet in order to be deemed trustworthy  \citenum{EU_2019_2}:
\begin{enumerate}[leftmargin=3mm,align=left,labelwidth=\parindent,labelsep=0pt]
\item Human agency and oversight: AI systems should empower human beings, allowing them to make informed decisions and fostering their fundamental rights. At the same time, proper oversight mechanisms need to be ensured, which can be achieved through: i) human-in-the-loop; ii) human-on-the-loop, and; iii) human-in-command approaches;
\item Technical Robustness and safety: AI systems need to be resilient and secure. They need to be safe, ensuring a fall back plan in case something goes wrong, as well as being accurate, reliable and reproducible. That is the only way to ensure that also unintentional harm can be minimized and prevented;
\item Privacy and data governance: besides ensuring full respect for privacy and data protection, adequate data governance mechanisms must also be ensured, taking into account the quality and integrity of the data, and ensuring legitimized access to data;
\item Transparency: the data, system and AI business models should be transparent. Traceability mechanisms can help achieving this. Moreover, AI systems and their decisions should be explained in a manner adapted to the stakeholder concerned. Humans need to be aware that they are interacting with an AI system, and must be informed of the system's capabilities and limitations;
\item Diversity, non-discrimination and fairness: Unfair bias must be avoided, as it could could have multiple negative implications, from the marginalization of vulnerable groups, to the exacerbation of prejudice and discrimination. Fostering diversity, AI systems should be accessible to all, regardless of any disability, and involve relevant stakeholders throughout their entire life circle;
\item Societal and environmental well-being: AI systems should benefit all human beings, including future generations. It must hence be ensured that they are sustainable and environmentally friendly. Moreover, they should take into account the environment, including other living beings, and their social and societal impact should be carefully considered;
\item Accountability: Mechanisms should be put in place to ensure responsibility and accountability for AI systems and their outcomes; 
\item Auditability, which enables the assessment of algorithms, data and design processes plays a key role therein, especially in critical applications. Moreover, adequate an accessible redress should be ensured.
\end{enumerate}

\subsection{Definition of non-performing loans and their subcategories by the Bank of Italy}
The definitions of non-performing loans (NPLs)  and their subcategories adopted by the Bank of Italy have been harmonized within the Single Supervisory Mechanism (SSM) and meet the European Banking Authority (EBA) standards \citenum{ECB_2017}, \citenum{ECB_2021},\citenum{BI_2008}, \citenum{BI_2020_1}, \citenum{BI_2020_2}, \citenum{BI_2021_1}, \citenum{BI_2021_2}, \citenum{BI_2021_3}, \citenum{BI_2022_1}.\\
NPLs are exposures to debtors who are no longer able to meet all or part of their contractual obligations because their economic and financial circumstances have deteriorated.\\
The 3 subcategories of NPLs are:
\begin{enumerate}[leftmargin=3mm,align=left,labelwidth=\parindent,labelsep=2pt]
\item Bad loans are exposures to debtors that are insolvent or in substantially similar circumstances.
\item Unlikely-to-pay exposures (aside from those included among bad loans) are those in respect of which banks believe the debtors are unlikely to meet their contractual obligations in full unless action such as the enforcement of guarantees is taken.
\item Overdrawn and/or past-due exposures (aside from those classified among bad loans and unlikely-to-pay exposures) are those that are overdrawn and/or past-due by more than 90 days and for above a predefined amount.
\end{enumerate}

\section{Methods}
\label{sec:Methods}

\subsection{Purpose of the 5D research project}
\label{sec:Purpose}
The goal of our research was to test if 5D, a machine-learning-based loan screening application, could detect a sub-subset of bad loans: 1) in a small subset of non-performing loans (bad loans, unlikely-to-pay, overdrawn and/or past-due exposures), and; 2) in a larger set of performing and non-performing loans subject to the following constraints: a) achieve a sensitivity and specificity not lower than 0.9: b) utilize a maximally relevant, minimally redundant number of publicly available features unrelated to the credit history, gender, race or ethnicity of the borrower (BiMOPT features); 3) comply with the European Banking Authority's and EU AI HLEG's Commission's principles on Big Data \& Advanced Analytics based on Artificial Intelligence.

\subsection{Phases}
\label{sec:Phases}
\begin{figure}[!h]
\centering
\begin{tikzpicture}[node distance=2cm, thick]
\node (start) [startstop] {5D Research Project};
\node (in0) [io, below of=start] {Phase 0: Bias-mitigated minimal-optimal features selection};
\node (in1) [io, below of=in0] {Phase I: Retrospective Training-Testing};
\node (pro1) [io, below of=in1] {Phase II: Baseline Validation};
\node (dec1) [io, below of=pro1] {Phase III: Prospective Real World Validation};
\draw [arrow] (start) -- (in0);
\draw [arrow] (in0) -- (in1);
\draw [arrow] (in1) -- (pro1);
\draw [arrow] (pro1) -- (dec1);
\end{tikzpicture}
\caption{Phases of the 5D Research Project \ref{sec:Phases}}
\label{fig:fig1}
\end{figure}
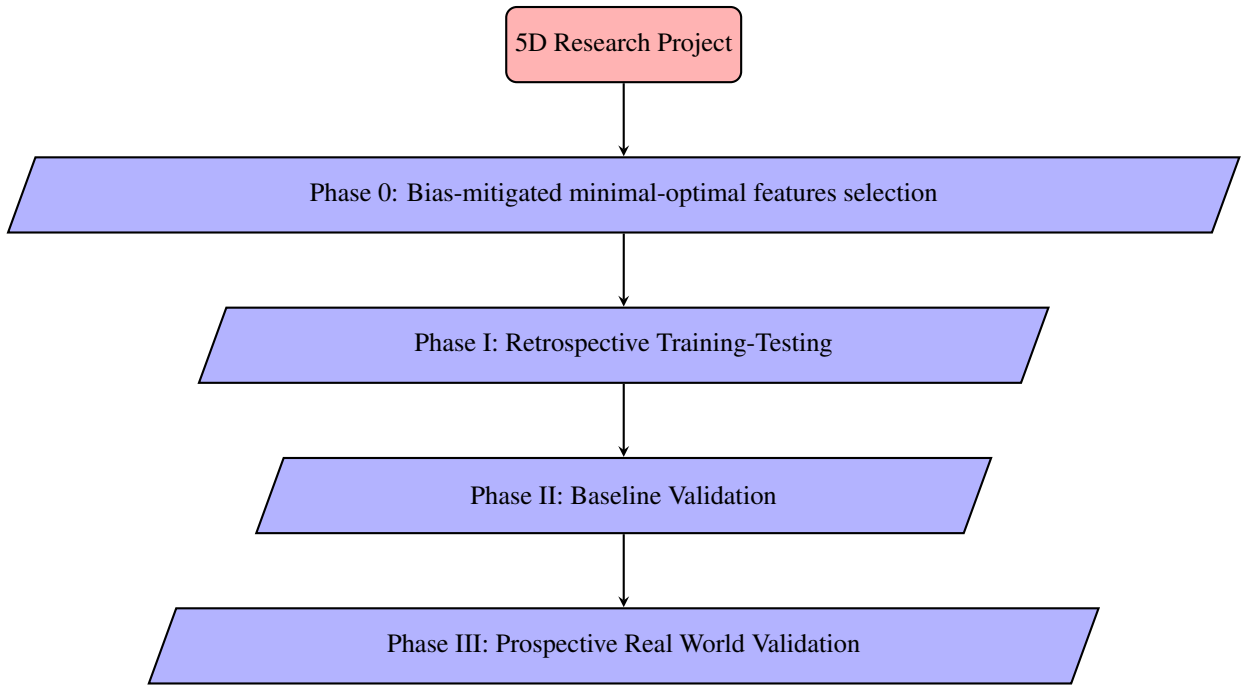

The 5D research project is divided into 4 Phases as shown in Figure \ref{fig:fig1}: 
\begin{enumerate}[leftmargin=3mm,align=left,labelwidth=\parindent,labelsep=2pt]
\item in Phase 0 we selected a subset of 10 BiMOPT features out of a total of 84 features;
\item in Phase I we trained 5D to detect bad loans in a historical dataset extracted from a mandatory report to the Bank of Italy consisting of 7,289 non-performing loans (NPLs) closed in the period 2010-2021;
\item in Phase II we assessed the baseline performance of 5D on a distinct validation dataset consisting of an active portolio of 63,763 outstanding loans (performing and non-performing) for a total financed value of over EUR 11.5 billion as of December 31, 2021; 
\item in Phase III we will monitor the baseline performance for a period of 5 years (2023-27) to assess the prospective real-world bias-mitigation and performance of the 5D system and its utility in credit and fintech institutions.
\end{enumerate}

\subsection{Timeline}
\label{sec:Timeline}
\begin{figure}[!h]
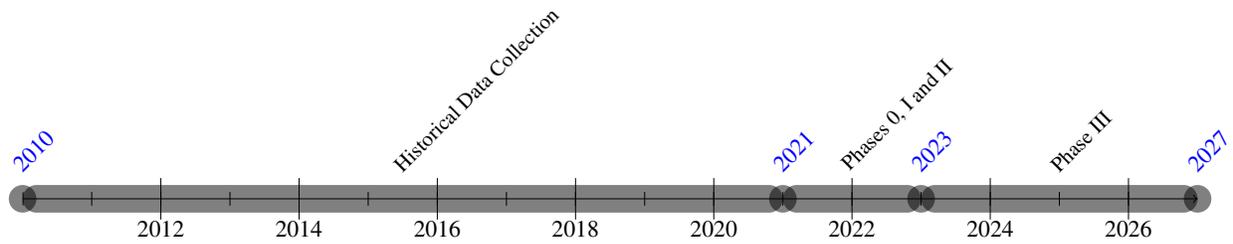

\centering
\begin{chronology}[2]{2010}{2026}{\textwidth}[\textwidth]
\event{2010}{\color{blue}{2010}}
\event[2010]{2021}{\small{Historical Data Collection}}
\event{2021}{\color{blue}{2021}}
\event[2021]{2023}{\small{Phases 0, I and II}}
\event{2023}{\color{blue}{2023}}
\event[2023]{2027}{\small{Phase III}}
\event{2027}{\color{blue}{2027}}
\end{chronology}
\caption{Timeline of the 5D research project \ref{sec:Timeline}}
\label{fig:fig2}
\end{figure}

The timeline of the 5D research project is shown in Figure \ref{fig:fig2}:

\subsection{Datasets}
\label{sec:Datasets}
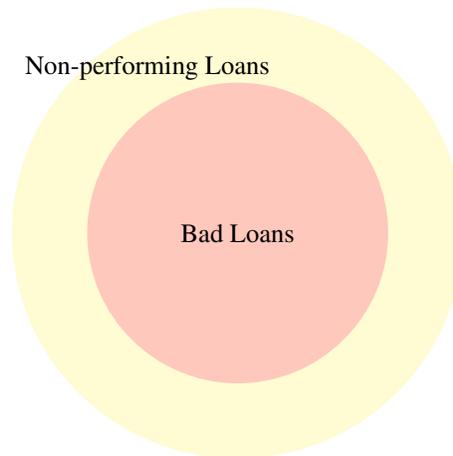
\begin{figure}[!h]
\centering
\begin{tikzpicture}
\begin{scope} [fill opacity = .7]
\fill[yellow!30!white]   ( 0:1.2) circle (3);
\fill[red!30!white] (0:1.2) circle (2);
\end{scope}
\node at ( 90:2.2)    {Non-performing Loans};
\node at (0:1.2)    {Bad Loans};
\end{tikzpicture}
\caption{Train-Test Dataset}
\label{fig:fig3}
\paragraph{Legend}
\raggedright
~\\
Composition of the Train-Test Dataset consisting only of Non-performing Loans (NPLs) closed in the period 2010-2021: i) NPLs are exposures to debtors who are no longer able to meet all or part of their contractual obligations because their economic and financial circumstances have deteriorated; ii) Bad loans is the most severe subcategory of NPLs consisting of exposures to debtors that are insolvent or in substantially similar circumstances \ref{sec:Datasets}.
\end{figure}

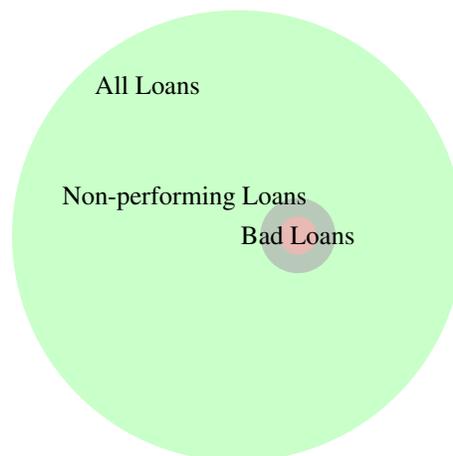
\begin{figure}[!h]
\centering
\begin{tikzpicture}
\begin{scope} [fill opacity = .7]
\fill[green!30!white]   ( 0:1.2) circle (3);
\fill[black!30!white] (0:2) circle (0.5);
\fill[red!30!white] (0:2) circle (0.25);
\end{scope}
\node at ( 90:2)    {All Loans};
\node at ( 45:.7)    {Non-performing Loans};
\node at (0:2)    {Bad Loans};
\end{tikzpicture}
\caption{Baseline Validation Dataset}
\label{fig:fig4}
\paragraph{Legend}
\raggedright
~\\
Composition of the Validation Dataset consisting of both Performing and Non-performing Loans (NPLs) as of December 31, 2021: i) NPLs are exposures to debtors who are no longer able to meet all or part of their contractual obligations because their economic and financial circumstances have deteriorated; ii) Bad loans is the most severe subcategory of NPLs consisting of exposures to debtors that are insolvent or in substantially similar circumstances \ref{sec:Datasets}.
\end{figure}

\begin{figure}[!h]
\centering
\includegraphics[width=150mm]{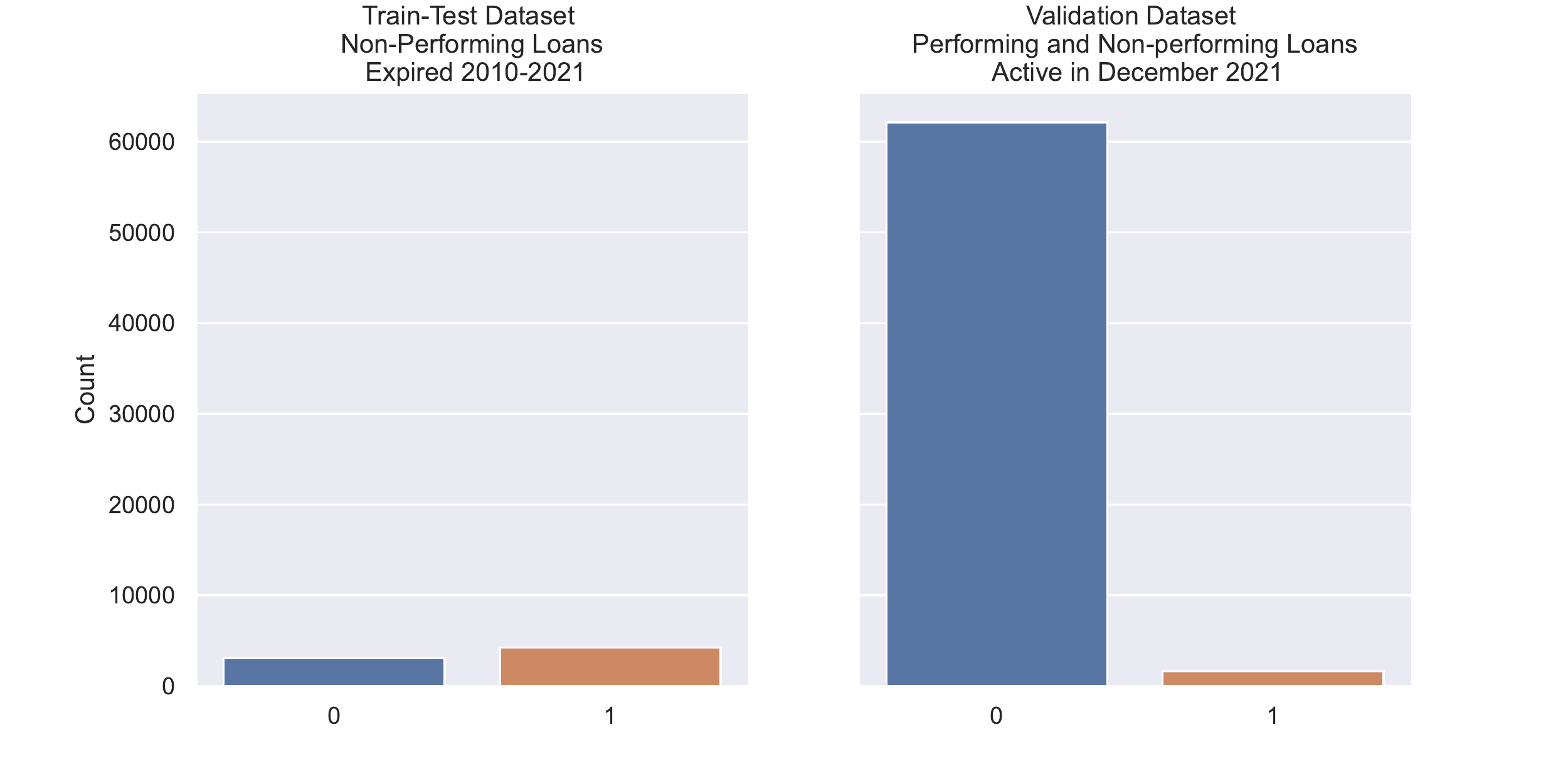}
\caption{Train-Test  and Baseline Validation Datasets}
\label{fig:fig5}
\paragraph{Legend}
\raggedright
~\\
Composition of the Train-Test Dataset \ref{sec:Datasets} consisting of:
\begin{itemize}[leftmargin=3mm,align=left,labelwidth=\parindent,labelsep=2pt]
\item  0 = Unlikely-to-pay exposures (aside from those included among bad loans) are those in respect of which banks believe the debtors are unlikely to meet their contractual obligations in full unless action such as the enforcement of guarantees is taken; 
\item 1 = Bad loans is the most severe subcategory of NPLs consisting of exposures to debtors that are insolvent or in substantially similar circumstances.
\end{itemize}
\raggedright
Composition of the Baseline Validation Dataset consisting of:
\begin{itemize}[leftmargin=3mm,align=left,labelwidth=\parindent,labelsep=2pt]
\item  0 = Performing and Non-performing exposures which are not Bad Loans; 
\item 1 = Bad loans.
\end{itemize}
\end{figure}

The 5D model has been trained and tested on the historical portfolio extracted from a mandatory report to the Bank of Italy  and validated on the active portfolio:
\begin{itemize}[leftmargin=3mm,align=left,labelwidth=\parindent,labelsep=2pt]
\item The historical portfolio extracted from a mandatory report to the Bank of Italy consists only of Non-performing Loans (NPLs) closed in the period 2010-2021 as shown in Figure \ref{fig:fig3}: i) NPLs are exposures to debtors who are no longer able to meet all or part of their contractual obligations because their economic and financial circumstances have deteriorated; ii) Bad loans is the most severe subcategory of NPLs consisting of exposures to debtors that are insolvent or in substantially similar circumstances;
\item The active portfolio consists of both Performing and Non-performing Loans (NPLs) as of December 31, 2021 as shown in Figure \ref{fig:fig4}.
\end{itemize}
The  historical portfolio (Train-Test Dataset) was balanced versus a highly imbalanced active portfolio (Baseline Validation Dataset) as shown in Figure \ref{fig:fig5} and Table \ref{tab:tab1}.

\subsection{EU GDPR Data Anonymization and Pseudonymization}
\label{sec:Anonymization}
All datasets have been anonymized and pseudoanonymized. We removed any references to an identifiable person/entity from all data sets, thus turning personal information into non-personal information. The remaining data have been pseudonymized by masking and noising, so that no person/entity can be identified from the  information without reference to additional information. The additional information has been kept separately from the pseudonymized person/entity information and is subject to technical and organizational safeguards (such as access controls) to keep it secure \citenum{EUdataregulations2018}, \citenum{katirai2006theory}. 

\subsection{ Bias-mitigated minimal-optimal features}
\label{sec:Features}
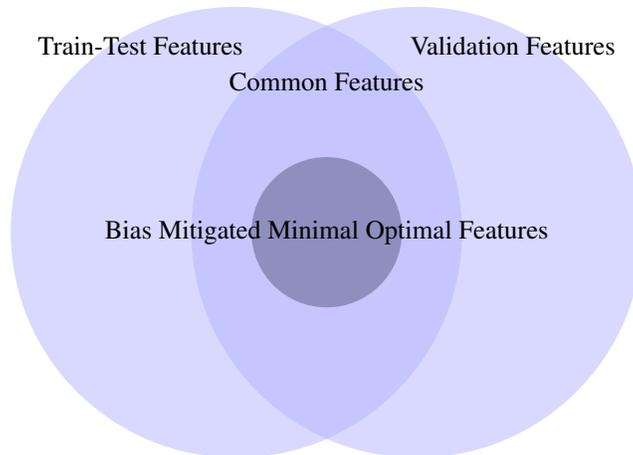
\begin{figure}[!h]
\centering
\begin{tikzpicture}
\begin{scope} [fill opacity = .5]
\fill[blue!30!white]   ( 0:1.2) circle (3);
 \fill[black!] (0:0) circle (1);
  \fill[blue!30!white] (180:1.2) circle (3);
 \end{scope}
 \node at ( 135:3.5)    {Train-Test Features};
 \node at (45:3.5)    {Validation Features};
\node at (90:2) {Common Features};
\node at (0:0) {Bias Mitigated Minimal Optimal Features};
\end{tikzpicture}
\caption{Bias-mitigated minimal-optimal features}
\label{fig:fig6}
\paragraph{Legend}
\raggedright
~\\
Bias-mitigated minimal-optimal features (BiMOPT) are bias-mitigated, maximum relevance, minimum redundance, publicly available features unrelated to the credit history, gender, race or ethnicity of the borrower. They are a subset of the features in common between the Train-Test and Baseline Validation Datasets \ref{sec:Features}.
\end{figure}

We performed data-based bias mitigation by \citenum{MIT_2020_1}, \citenum{MIT_2020_2}, \citenum{zhong2018tutorial}, \citenum{olteanu2018tutorial}, \citenum{kun2015one} removing features related to the credit history, gender, race or ethnicity of the borrower \citenum{verma2021removing}, \citenum{abu2020contextual} .\\
We improved the performance of 5D by selecting maximally relevant, minimally redundant features \citenum{cover1974best}, \citenum{ding2005minimum}, \citenum{zhao2019maximum}.\\
We referred to these new features as Bias-Mitigated Minimal-Optimal Features (BiMOPT) as shown in Figure \ref{fig:fig6} and Table \ref{tab:tab2} .

\subsection{Retrospective Performance of the 5D model on the Train-Test Dataset}
\label{sec:Retrospective}
\begin{figure}[t!]
\centering
\includegraphics[width=150mm]{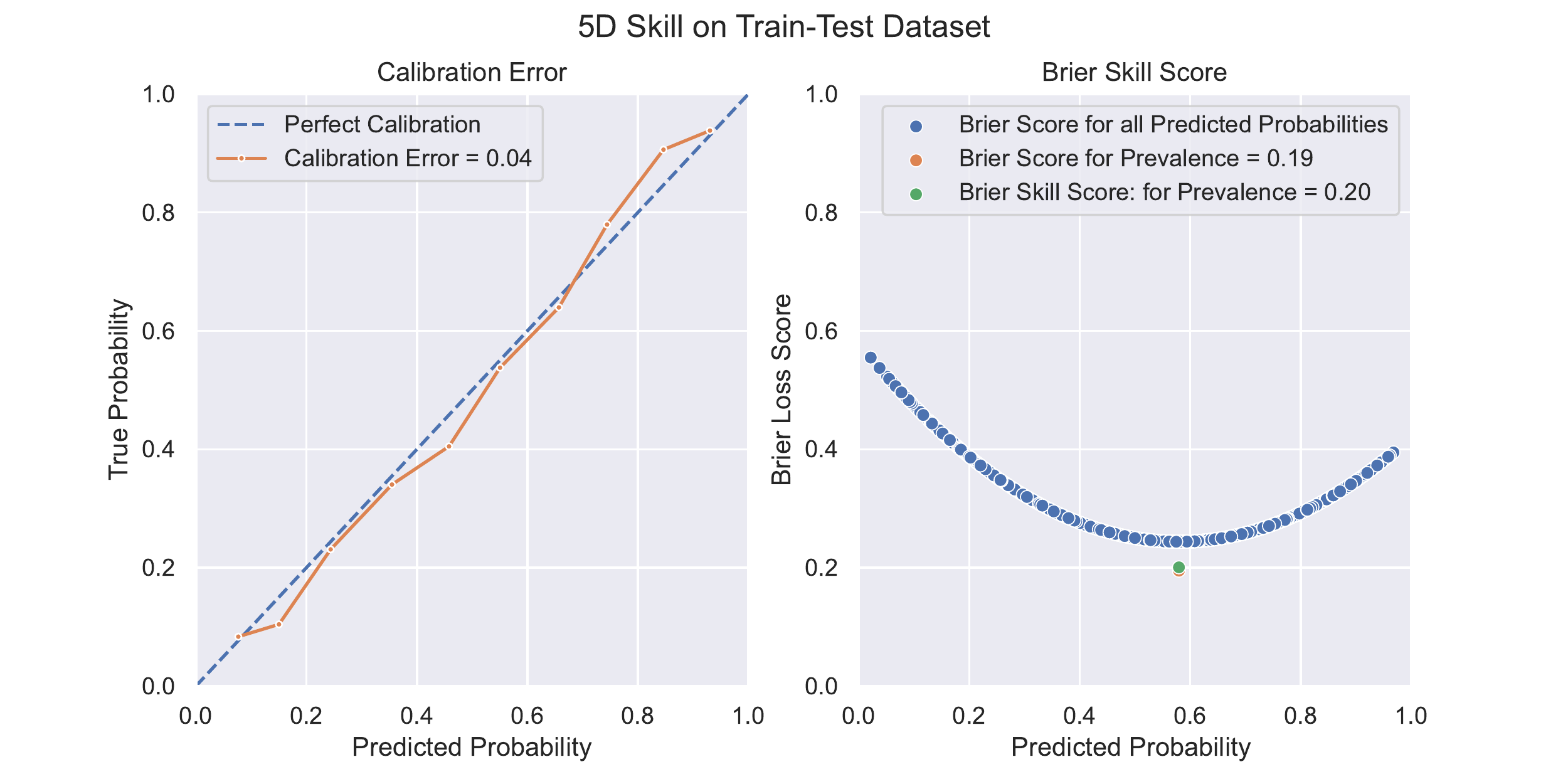}
\caption{Retrospective performance of the 5D model on the train-test dataset}
\label{fig:fig7}
\paragraph{Legend}
\raggedright
~\\
Calibration, Brier Loss Score and  Brier Skill Score of 5D in the training-testing retrospective Phase II \ref{sec:Retrospective}:
\begin{enumerate}[leftmargin=3mm,align=left,labelwidth=\parindent,labelsep=2pt]
\item The Calibration Curve (reliability diagram) compares how well the probabilistic predictions of the 5D binary classifier are calibrated. It plots the true frequency of the positive label (Bad Loans) against its predicted probability, for binned predictions. The x axis represents the average predicted probability in each bin. The y axis is the fraction of positives, i.e. the proportion of samples whose class is the positive class in each bin (Bad Loans); \\
\item Thus, Expected Calibration Error (ECE) is a weighted mean of the calibration errors of the single bins, where each bin weighs proportionally to the number of observations that it contains. We set the number of bins according to the Freedman-Diaconis rule  designed for finding the number of bins that makes the histogram as close as possible to the theoretical probability distribution; \\
\item The Brier Loss Score  (BS) \ref{eq:1} measures the mean squared difference between the predicted probability and the actual outcome. The smaller the Brier score loss, the better, hence the naming with loss. The Brier score always takes on a value between zero and one, since this is the largest possible difference between a predicted probability (which must be between zero and one) and the actual outcome (which can take on values of only 0 and 1). The Brier score is appropriate for binary and categorical outcomes. The x axis represents the predicted probability. The y axis is the Brier Loss Score of the classifier for every predicted probability;\\
\item The quality of the predictions and thus the reliability of probability predictions can be evaluated using the Brier Skill Score (BSS) \ref{eq:2}. The BSS is based on the Brier Score (BS). BSS  for a given underlying score is an offset and (negatively) scaled variant of the underlying score such that a skill score value of zero means that the score for the predictions is merely as good as that of a set of baseline or reference or default predictions, while a skill score value of one  represents the best possible score. A skill score value less than zero means that the performance is even worse than that of the baseline or reference predictions.
\end{enumerate}
\end{figure}

We trained a scalable tree-based extreme gradient boosting meta-algorithm \citenum{Chen:2016:XST:2939672.2939785} and assessed its performance and skill on the Train-Test Dataset by calibrating the predicted probabilities \citenum{niculescu2005predicting} and by measuring the Brier Loss  \citenum{brier1950verification} and Brier Skill Score (Figure \ref{fig:fig7})  \citenum{gneiting2007strictly}.
\begin{enumerate}[leftmargin=3mm,align=left,labelwidth=\parindent,labelsep=2pt]
\item The Expected Calibration Error (ECE) curve (reliability diagram) compares how well the probabilistic predictions of the 5D binary classifier are calibrated. It plots the true frequency of the positive label (Bad Loans) against its predicted probability, for binned predictions  \citenum{naeini2015obtaining},  \citenum{guo2017calibration},  \citenum{niculescu2005predicting};\\
\item The Brier Loss Score  (BS) measures the mean squared difference between the predicted probability and the actual outcome.
 The smaller the Brier score loss, the better, hence the naming with loss \ref{eq:1}. 
 
 \begin{equation}
 \label{eq:1}
 \begin{array}{l}
 BS=\frac{1}{N}\displaystyle \sum_{t=1}^{N} (f_t-o_t)^2 ~~~~~~ (0 \le BS \le 1)
 \end{array}
\end{equation}

 The Brier Score always takes on a value between zero and one, since this is the largest possible difference between a predicted probability (which must be between zero and one) and the actual outcome (which can take on values of only 0 and 1). The Brier score is appropriate for binary and categorical outcomes \citenum{brier1950verification};\\
\item The quality of the predictions and thus the reliability of probability predictions can be evaluated using the Brier Skill Score (BSS) \ref{eq:2}.

\begin{equation}
\label{eq:2}
 \begin{array}{l}
 BSS=1-\frac{BS}{BS_{ref}}  ~~~~~~ (BSS \le 1)
 \end{array}
\end{equation} 

Where BS is the Brier Score, BS\textsubscript{ref} is the No-skill Brier Score, or the Brier Score resulting by just predicting the positive class (bad loans) based on prevalence, in this case 0.0253. \\
The BSS indicates the degree of improvement of the classifier over a classifier with no skill. A skill score value less than zero means that the performance is even worse than that of the baseline or reference predictions \ref{tab:tab3}.
\end{enumerate}

\section{Results}
\label{sec:Results}
Using as little as 10 Bias-mitigated minimal-optimal features \ref{sec:Features} selected out of 84 total features, 5D correctly detected 1,461 bad loans out of a total of 1,613 (Sensitivity = 0.91, Prevalence = 0.0253), and correctly classified 55,866 out of the other 62,150 exposures (Specificity = 0.90) \ref{tab:tab4}.\\
The low Positive Predictive Value (0.19) and high Negative Predictive Value (0.997) of bad loans by 5D at baseline is characteristic of a 'prudential' approach to the classification of bad loans, as banks are required to adopt 'prudential' credit risk models. Only in Phase III of the research project we will be able to update the baseline performances of the 5D system.

\section{Limitations}
\label{sec:Limitations}
To our present knowledge, our research findings suffer from the following limitations:
\begin{itemize}[leftmargin=3mm, align=left,labelwidth=\parindent,labelsep=2pt]
\item 5D has been baseline validated on an 'active' portfolio. This means that the baseline performance values reported in the \ref{sec:Results} section can be 'terminally' estimated only when the loans they refer to are closed;
\item the 5D classifier has been trained on the portfolio of a specific bank-related asset financing institution. This means that it might not 'generalize' to other credit institutions, even if trained on their specific datasets;
\item the 10 BiMOPT features selected for the machine-learning-based predictions can be time-biased; in other words, they might contain information which has been updated in the period 2010-2021 to reflect the evolving status of the loan; that information would not have been available at the time of the decision, therefore 5D has an 'advantage' towards an original 'manual' decisor;
\item the 10 BiMOPT features might be subject to 'data drift' bot in the period 2010-2021 and in the Phase III period 2022-2027;  drift is a distribution change between the training and deployment data, which can affect model performance for reasons not related to the actual predictive power of the algorithm \citenum{ackerman2020detection}, \citenum{ackerman2021automatically}. 
\end{itemize}

\section{Interpretation}
\label{sec:Interpretation}
Our preliminary results support the hypothesis that Big Data \& Advanced Analytics applications based on AI can mitigate bias and improve consumer protection in the loan screening process without compromising the efficacy of the credit risk assessment. Further validation is required; in Phase III of the research project we will monitor the baseline performance of 5D for a period of 5 years (2023-27) to assess the prospective real-world bias-mitigation and performance of the 5D system and its utility in credit and fintech institutions.

\section{Funding}
This study received  funding from QuantumSPEKTRAL (Milano, Italy, EU) and Alba Leasing (Milano, Italy, EU) in the form of direct reimbursement of research costs. No other funding was received.

\section*{Acknowledgments}
We are grateful of the collaboration on use cases and empirical evaluation by the employees of Alba Leasing. In particular, we appreciate Stefano Rossi for the guidance and support on implementation. We also would like to extend our gratitude to Nima Heidari and Brady Fish for for reviewing the manuscript. We would also like to thank a number of anonymous reviewers for helpful comments on earlier versions of this paper.

\section*{Author contributions statement}
S.O. and D.M. were responsible for conceptualization, the methodology, and the software; A.D. and S.O. were responsible for the formal analysis; S.O. and M.R. were responsible for the investigation; A.D. was responsible for the resources; M.R. was responsible for data curation; S.O. was responsible for writing the original draft preparation; A.D., M.R. and F.M. were responsible for review and editing; S.O. and D.M. were responsible for visualization; A.D., M.R. and F.M. were responsible for supervision; A.D. was responsible for project administration. S.O. and D.M. are responsible for the AI/ML algorithms. All authors have read and agreed to the published version of the manuscript.

\section*{Competing Interest Statement}
The authors declare that they have received support for the present manuscript in terms of provision of funds, data and study materials. In the past 36 months, the authors have received grants from public research institutions and contracts from fintech and other financial companies pertaining the field of fintech. The authors have received and are likely to receive in the future royalties, licenses, consulting fees, payment or honoraria for lectures, presentations, speakers bureaus, manuscript writing or educational events. Some parts of the Artificial Intelligence algorithms are copyrighted by QuantumSPEKTRAL and QuantumSPEKTRAL is likely to patent some of the contents. The authors hold stock and stock options directly and indirectly in publicly traded and private fintech and other financial companies. The authors declare no other competing interests.

\section*{Copyright, Patents and Registered Marks}
The copyright holder for this preprint is QuantumSPEKTRAL  who has granted a license to display the preprint in perpetuity. All rights reserved. No reuse allowed without permission. 5D is a registered trademark of QuantumSPEKTRAL.

\section*{Machine Learning / Artificial Intelligence Disclaimer}
Machine Learning and Artificial Intelligence are sometimes wrong. Although QuantumSPEKTRAL (Milano, Italy) takes all possible care to ensure the correctness of published information, no warranty can be accepted regarding the correctness, accuracy, up-to-dateness, reliability and completeness of the content of this information. QuantumSPEKTRAL expressly reserves the right to change, delete or temporarily not to publish the contents of this preprint wholly or partly at any time and without giving notice. Liability claims against QuantumSPEKTRAL  because of tangible or intangible damage arising from accessing, using or not using the published information are excluded.

\section*{EU GDPR Data Anonymization and Pseudonymization Policy}
All datasets used to train, test, and validate the machine-learning models have been destroyed by QuantumSPEKTRAL after training, testing, and validation. Before training, testing, and validation, Alba Leasing and QuantumSPEKTRAL removed any references to an identifiable person/entity from all data sets, thus turning personal information into non-personal information. The remaining data have been pseudonymized by masking and noising, so that no person/entity can be identified from the  information without reference to additional information. The additional information has been kept separately from the pseudonymized person/entity information and is subject to technical and organizational safeguards (such as access controls) to keep it secure. Given the changing nature of technology, it is possible that some anonymized data sets might, one day, be subject to re-identification. Some researchers also found that machine-learning models can be used to re-build the original datasets. However, this should not be reasonably possible in the current technological climate.


\bibliographystyle{unsrtnat}
\bibliography{QSPK_Alba_3_3} 

\begin{thebibliography}{34}
\providecommand{\natexlab}[1]{#1}
\providecommand{\url}[1]{\texttt{#1}}
\expandafter\ifx\csname urlstyle\endcsname\relax
  \providecommand{\doi}[1]{doi: #1}\else
  \providecommand{\doi}{doi: \begingroup \urlstyle{rm}\Url}\fi

\bibitem[Authority(2020)]{EBA_2020}
European~Banking Authority.
\newblock Report on big data and advanced analytics, 2020.

\bibitem[Authority(2016)]{EBA_2016}
European~Banking Authority.
\newblock Joint committee discussion paper on the use of big data by financial
  institutions, 2016.

\bibitem[Authority(2018)]{EBA_2018}
European~Banking Authority.
\newblock Joint committee final report on big data, 2018.

\bibitem[on~Artificial~Intelligence(2019{\natexlab{a}})]{EU_2019_1}
European Commission High-Level Expert~Group on~Artificial~Intelligence.
\newblock Ethics guidelines for trustworthy ai, 2019{\natexlab{a}}.

\bibitem[on~Artificial~Intelligence(2019{\natexlab{b}})]{EU_2019_2}
European Commission High-Level Expert~Group on~Artificial~Intelligence.
\newblock A definition of ai: main capabilities and disciplines. definition
  developed for the purpose of the ai hleg deliverables, 2019{\natexlab{b}}.

\bibitem[Bank(2017)]{ECB_2017}
European~Central Bank.
\newblock Guidance to banks on non-performing loanss, 2017.

\bibitem[Bank(2021)]{ECB_2021}
European~Central Bank.
\newblock Non-performing loans. supervisory practices. priorities and risks,
  2021.

\bibitem[of~Italy(2008)]{BI_2008}
Bank of~Italy.
\newblock Circolare n. 272 del 30 luglio 2008 - matrice dei conti, 2008.

\bibitem[of~Italy(2020{\natexlab{a}})]{BI_2020_1}
Bank of~Italy.
\newblock Financial stability report no. 1 - 2020, 2020{\natexlab{a}}.

\bibitem[of~Italy(2020{\natexlab{b}})]{BI_2020_2}
Bank of~Italy.
\newblock Financial stability report no. 2 - 2020, 2020{\natexlab{b}}.

\bibitem[of~Italy(2021{\natexlab{a}})]{BI_2021_1}
Bank of~Italy.
\newblock Financial stability report no. 1 - 2021, 2021{\natexlab{a}}.

\bibitem[of~Italy(2021{\natexlab{b}})]{BI_2021_2}
Bank of~Italy.
\newblock Financial stability report no. 2 - 2021, 2021{\natexlab{b}}.

\bibitem[of~Italy(2021{\natexlab{c}})]{BI_2021_3}
Bank of~Italy.
\newblock Circolare n. 272 del 30 luglio 2008 - matrice dei conti - testo
  integrale al 15 aggiornamento, 2021{\natexlab{c}}.

\bibitem[of~Italy(2022)]{BI_2022_1}
Bank of~Italy.
\newblock Financial stability report no. 1 - 2022, 2022.

\bibitem[Commission(2018)]{EUdataregulations2018}
European Commission.
\newblock Eu general data protection regulation (gdpr), 2018.
\newblock URL
  \url{https://ec.europa.eu/commission/sites/beta-political/files/data-protection-factsheet-changes_en.pdf}.

\bibitem[Katirai(2006)]{katirai2006theory}
Hooman Katirai.
\newblock \emph{A theory and toolkit for the mathematics of privacy: methods
  for anonymizing data while minimizing information loss}.
\newblock PhD thesis, Massachusetts Institute of Technology, 2006.

\bibitem[Review(2020)]{MIT_2020_1}
MIT Sloan~Management Review.
\newblock The risk of machine learning bias (and how to prevent it), 2020.

\bibitem[of~Technology(2020)]{MIT_2020_2}
Massachusetts~Institute of~Technology.
\newblock Case study with data: Mitigating gender bias on the uci adult
  database, 2020.

\bibitem[Zhong(2018)]{zhong2018tutorial}
Ziyuan Zhong.
\newblock A tutorial on fairness in machine learning.
\newblock \emph{Medium, October}, 22:\penalty0 2018, 2018.

\bibitem[Olteanu(2018)]{olteanu2018tutorial}
A~Olteanu.
\newblock Tutorial: Learning curves for machine learning in python, 2018.

\bibitem[Kun(2015)]{kun2015one}
J~Kun.
\newblock One definition of algorithmic fairness: Statistical parity, 2015.

\bibitem[Verma et~al.(2021)Verma, Ernst, and Just]{verma2021removing}
Sahil Verma, Michael Ernst, and Rene Just.
\newblock Removing biased data to improve fairness and accuracy.
\newblock \emph{arXiv preprint arXiv:2102.03054}, 2021.

\bibitem[Abu-Elyounes(2020)]{abu2020contextual}
Doaa Abu-Elyounes.
\newblock Contextual fairness: A legal and policy analysis of algorithmic
  fairness.
\newblock \emph{U. Ill. JL Tech. \& Pol'y}, page~1, 2020.

\bibitem[Cover(1974)]{cover1974best}
Thomas~M Cover.
\newblock The best two independent measurements are not the two best.
\newblock \emph{IEEE Transactions on Systems, Man, and Cybernetics}, 1\penalty0
  (1):\penalty0 116--117, 1974.

\bibitem[Ding and Peng(2005)]{ding2005minimum}
Chris Ding and Hanchuan Peng.
\newblock Minimum redundancy feature selection from microarray gene expression
  data.
\newblock \emph{Journal of bioinformatics and computational biology},
  3\penalty0 (02):\penalty0 185--205, 2005.

\bibitem[Zhao et~al.(2019)Zhao, Anand, and Wang]{zhao2019maximum}
Zhenyu Zhao, Radhika Anand, and Mallory Wang.
\newblock Maximum relevance and minimum redundancy feature selection methods
  for a marketing machine learning platform.
\newblock In \emph{2019 IEEE International Conference on Data Science and
  Advanced Analytics (DSAA)}, pages 442--452. IEEE, 2019.

\bibitem[Chen and Guestrin(2016)]{Chen:2016:XST:2939672.2939785}
Tianqi Chen and Carlos Guestrin.
\newblock {XGBoost}: A scalable tree boosting system.
\newblock In \emph{Proceedings of the 22nd ACM SIGKDD International Conference
  on Knowledge Discovery and Data Mining}, KDD '16, pages 785--794, New York,
  NY, USA, 2016. ACM.
\newblock ISBN 978-1-4503-4232-2.
\newblock \doi{10.1145/2939672.2939785}.
\newblock URL \url{http://doi.acm.org/10.1145/2939672.2939785}.

\bibitem[Niculescu-Mizil and Caruana(2005)]{niculescu2005predicting}
Alexandru Niculescu-Mizil and Rich Caruana.
\newblock Predicting good probabilities with supervised learning.
\newblock In \emph{Proceedings of the 22nd international conference on Machine
  learning}, pages 625--632, 2005.

\bibitem[Brier et~al.(1950)]{brier1950verification}
Glenn~W Brier et~al.
\newblock Verification of forecasts expressed in terms of probability.
\newblock \emph{Monthly weather review}, 78\penalty0 (1):\penalty0 1--3, 1950.

\bibitem[Gneiting and Raftery(2007)]{gneiting2007strictly}
Tilmann Gneiting and Adrian~E Raftery.
\newblock Strictly proper scoring rules, prediction, and estimation.
\newblock \emph{Journal of the American statistical Association}, 102\penalty0
  (477):\penalty0 359--378, 2007.

\bibitem[Naeini et~al.(2015)Naeini, Cooper, and
  Hauskrecht]{naeini2015obtaining}
Mahdi~Pakdaman Naeini, Gregory Cooper, and Milos Hauskrecht.
\newblock Obtaining well calibrated probabilities using bayesian binning.
\newblock In \emph{Twenty-Ninth AAAI Conference on Artificial Intelligence},
  2015.

\bibitem[Guo et~al.(2017)Guo, Pleiss, Sun, and Weinberger]{guo2017calibration}
Chuan Guo, Geoff Pleiss, Yu~Sun, and Kilian~Q Weinberger.
\newblock On calibration of modern neural networks.
\newblock In \emph{International conference on machine learning}, pages
  1321--1330. PMLR, 2017.

\bibitem[Ackerman et~al.(2020)Ackerman, Farchi, Raz, Zalmanovici, and
  Dube]{ackerman2020detection}
Samuel Ackerman, Eitan Farchi, Orna Raz, Marcel Zalmanovici, and Parijat Dube.
\newblock Detection of data drift and outliers affecting machine learning model
  performance over time.
\newblock \emph{arXiv preprint arXiv:2012.09258}, 2020.

\bibitem[Ackerman et~al.(2021)Ackerman, Raz, Zalmanovici, and
  Zlotnick]{ackerman2021automatically}
Samuel Ackerman, Orna Raz, Marcel Zalmanovici, and Aviad Zlotnick.
\newblock Automatically detecting data drift in machine learning classifiers.
\newblock \emph{arXiv preprint arXiv:2111.05672}, 2021.

\end{thebibliography}

\end{multicols}
\newpage

\section{Tables}
\label{sec:Tables}

\begin{table}[!h]
\centering
\caption{Distribution of Bad Loans in Train-Test and Validation Datasets}
\label{tab:tab1}
\begin{tabular}{@{}|l|l|l|l|l|@{}}
\toprule
                   & Total & Bad Loans & Bad Loans as \% of Total & Bad Loans Distribution \\ \midrule
Train-Test Dataset & 7,289  & 4,224      & 57.95                    & Balanced               \\ \midrule
Validation Dataset & 63,763 & 1,613      & 2.53                     & Highly Imbalanced      \\ \bottomrule
\end{tabular}
\paragraph{Legend}
\raggedright
~\\
Composition of the Train-Test and Baseline Validation Datasets. The Baseline Validation Dataset is heavily imbalanced  \ref{sec:Datasets}.
\end{table}

\begin{table}[!h]
\centering
\caption{Bias-mitigated minimal-optimal features in Train-Test and Baseline Validation Datasets}
\label{tab:tab2}
\begin{tabular}{@{}|l|l|l|l|@{}}
\toprule
 & \begin{tabular}[c]{@{}l@{}}In the Train-Test \\ Dataset\end{tabular} & \begin{tabular}[c]{@{}l@{}}In the Baseline Validation \\ Dataset\end{tabular} & BiMOPT Features \\ \midrule
Number of Features & 30                                                                   & 84                                                                            & 10              \\ \bottomrule
\end{tabular}
\paragraph{Legend}
\raggedright
~\\
Bias-mitigated minimal-optimal features (BiMOPT) are bias-mitigated, maximum relevance, minimum redundance, publicly available features unrelated to the credit history, gender, race or ethnicity of the borrower. They are a subset of the features in common between the Train-Test and Baseline Validation Datasets \ref{sec:Features}.
\end{table}

\begin{table}[!h]
\centering
\caption{Bias-mitigated minimal-optimal features in Train-Test and Baseline Validation Datasets}
\label{tab:tab3}
\begin{tabular}{@{}|l|l|l|l|@{}}
\toprule
                       & ECE  & BS   & BSS \\ \midrule
Train-Test Performance & 0.04 & 0.19 & 0.2 \\ \bottomrule
\end{tabular}
\paragraph{Legend}
\raggedright
~\\
Calibration, Brier Loss Score and  Brier Skill Score of 5D in the training-testing retrospective Phase II \ref{sec:Retrospective}:
\begin{enumerate}[leftmargin=3mm,align=left,labelwidth=\parindent,labelsep=2pt]
\item Thus, Expected Calibration Error (ECE) is a weighted mean of the calibration errors of the single bins, where each bin weighs proportionally to the number of observations that it contains. We set the number of bins according to the Freedman-Diaconis rule  designed for finding the number of bins that makes the histogram as close as possible to the theoretical probability distribution; \\
\item The Brier Loss Score  (BS) \ref{eq:1} measures the mean squared difference between the predicted probability and the actual outcome. The smaller the Brier score loss, the better, hence the naming with loss. The Brier score always takes on a value between zero and one, since this is the largest possible difference between a predicted probability (which must be between zero and one) and the actual outcome (which can take on values of only 0 and 1). The Brier score is appropriate for binary and categorical outcomes. The x axis represents the predicted probability. The y axis is the Brier Loss Score of the classifier for every predicted probability;\\
\item The quality of the predictions and thus the reliability of probability predictions can be evaluated using the Brier Skill Score (BSS) \ref{eq:2}. The BSS is based on the Brier Score (BS). BSS  for a given underlying score is an offset and (negatively) scaled variant of the underlying score such that a skill score value of zero means that the score for the predictions is merely as good as that of a set of baseline or reference or default predictions, while a skill score value of one  represents the best possible score. A skill score value less than zero means that the performance is even worse than that of the baseline or reference predictions.
\end{enumerate}.
\end{table}

\begin{table}[!t]
\centering
\caption{5D Baseline Performance on the Validation Dataset}
\label{tab:tab4}
\begin{tabular}{@{}|l|l|l|l|l|l|@{}}
\toprule
                            & Prevalence of Bad Loans & Sensitivity & Specificity & PPV & NPV \\ \midrule
Baseline Performance & 0.0253                   & 0.90        & 0.91        & 0.19                      & 0.997                     \\ \bottomrule
\end{tabular}
\paragraph{Legend}
\raggedright
~\\
Baseline Performance of 5D \ref{sec:Results}:
\begin{enumerate}[leftmargin=3mm,align=left,labelwidth=\parindent,labelsep=2pt]
\item  Prevalence: Serves to measure the balance of data within the total population. It is possible to measure the prevalence of positives or negatives and the sum of both quotients is = 1, a balanced data set would give coefficients close to 0.5 If, on the contrary, one of the factors is close to 1 and the other to 0, we are going to have an unbalanced data set;
\item Sensitivity measures the proportion of true positives that are correctly identified as such; 
\item Specificity (also called the True Negative Rate) measures the proportion of true negatives that are correctly identified as such;\item  PPV - Positive Predictive Value or Precision describe the performance of a diagnostic test as the proportion of true positives in the predicted positives. PPV is not intrinsic to the test; it depends also on the Prevalence;
\item NPV - Negative predictive value describe the performance of a diagnostic test as the proportion of true negatives in the predicted negatives. NPV is not intrinsic to the test; it depends also on the Prevalence.
\end{enumerate}
\end{table}

\clearpage

\section{Figures}
\label{sec:Figures}

\vspace{1cm}

\vspace{1cm}

\vspace{2cm}

\end{document}